\documentclass[aip,jcp,reprint,showpacs,amsmath,amssymb,amsfonts,floatfix]{revtex4-1}
\usepackage{bm}
\usepackage{graphicx}

\usepackage[bookmarks,bookmarksopen,bookmarksnumbered,colorlinks,linkcolor=red,linktocpage,
citecolor=blue,urlcolor=cyan,pdfpagemode=UseOutline]{hyperref}

\newcommand{\br}{{\bf r}}
\newcommand{\parref}[1]{(\ref{#1})}
\newcommand{\ext}{_{\rm ext}}
\newcommand{\sss}{\scriptscriptstyle\rm}
\newcommand{\xc}{_{\sss XC}}
\newcommand{\Hxc}{_{\sss HXC}}
\newcommand{\vect}[1]{\mathbf{#1}}
\newcommand{\intd}{\mathrm{d}}
\providecommand{\abs}[1]{\left|#1\right|}
\newcommand{\matelem}[3]{\left\langle #1 \left| #2 \right| #3 \right\rangle}

\begin{document}
\title{A minimal model for excitons within time-dependent density-functional theory}
\author{Zeng-hui Yang, Yonghui Li, and Carsten A. Ullrich}
\affiliation{Department of Physics and Astronomy, University of Missouri, Columbia, MO 65211}
\date{\today}

\begin{abstract}
The accurate description of the optical spectra of insulators and semiconductors remains an important challenge for time-dependent
density-functional theory (TDDFT). Evidence has been given in the literature that TDDFT can produce bound as well
as continuum excitons for specific systems, but there are still many unresolved basic questions concerning the role of dynamical exchange and
correlation (xc). In particular, the role of the long spatial range and the frequency dependence of the xc kernel $f_{\rm xc}$
for excitonic binding are still not very well explored. We present a minimal model for excitons in TDDFT, consisting of two bands from
a one-dimensional Kronig-Penney model and simple approximate xc kernels, which allows us to
address these questions in a transparent manner. Depending on the system, it is found that adiabatic xc kernels can produce a single
bound exciton, and sometimes two bound excitons, where the long spatial range of $f_{\rm xc}$ is not a necessary condition. It is shown
how the Wannier model, featuring an effective electron-hole interaction, emerges from TDDFT. The collective, many-body nature of excitons is
explicitly demonstrated.
\end{abstract}

\maketitle

\section{Introduction}
\label{sect:intro}
The study of the electronic structure of materials usually begins
with noninteracting electrons due to the vast number of particles involved.
Many-body methods then provide a hierarchy of corrections to account for
the Coulomb interaction to various order. Many-body approaches such as
GW\cite{H65,AG98} and the Bethe-Salpeter equation (BSE)\cite{HS80,ORR02} are frequently and successfully employed in the
calculation of the electronic structure and excitations of materials.  Though
accurate and physically sound, these many-body methods can become
cumbersome and impractical for large systems due to the
steep scaling of the numerical cost versus the system size.

Alternatively, density-functional theory (DFT) and time-dependent density-functional
theory (TDDFT)\cite{RG84,MUNR06,U12} are popular methods for calculating electronic ground states and excitations,
respectively, and are widely used in chemistry, physics, materials science, and other areas. Density-functional methods
solve the many-body problem by constructing a noninteracting system which
reproduces the electronic density of the interacting,
physical system. The favorable balance between accuracy and efficiency makes the resulting DFT and TDDFT schemes
unrivaled for large but finite system sizes.\cite{EFB09}

Considerable effort has been spent to replicate this success of TDDFT for periodic solids.\cite{BSDR07}
Generally speaking, TDDFT works very well for simple metallic systems, where the excitation spectrum is
dominated by collective plasmon modes. The reason is that common local and semilocal exchange-correlation (xc) functionals
are based on the homogeneous electron liquid as reference system, which is an ideal starting point to
describe electrons in metals.

The situation is more complicated in insulators and semiconductors. The first problem that comes to mind
is that of the band gap, which is typically strongly underestimated by most popular xc functionals of DFT.
In principle, TDDFT provides a mechanism to obtain the correct band gap,\cite{GS99,STP04,GORT07}
but this puts very strong demands on the xc kernel $f_{\rm xc}$ (it has to simulate a discontinuity, which requires a strong frequency
dependence).

The second difficulty are excitonic effects. It is a well-known fact that standard local and semilocal xc functionals
do not produce any excitonic binding;\cite{ORR02,BSDR07} again, the proper choice of $f_{\rm xc}$ is crucial.
There are many examples in the literature of successful TDDFT calculations of excitonic effects, using
exact exchange, \cite{KG02} an effective xc kernel engineered from the BSE,\cite{RORO02,SKRA03,MDR03,DAOR03,SMOR07}
a meta-GGA kernel,\cite{NV11} and a recent `bootstrap' xc kernel.\cite{SDSG11} These kernels all have in common
that they have a long spatial range; however, it has also been shown that certain excitonic features can be
equally well reproduced by simple short-range kernels. \cite{SKRA03,BSDR07} This calls for further explanation.

Due to the complexity of real solids, the question of the general requirements for excitonic binding in TDDFT
has been difficult to analyze. As a first step towards a simplified TDDFT approach for excitons,
a two-band model was recently developed, which was used to test the performance of simple xc kernels for calculating excitonic
binding energies in several III-V and II-VI semiconductors.\cite{TU08,TLU09}
In this paper we will push this reductionist approach further and propose a minimal TDDFT model for excitons.

Our model is one-dimensional (1D) and uses two simple Kronig-Penney-type bands as input.
We show that the minimal model reproduces and reveals many aspects
of excitons. The model is accessible and relatively easy to implement,
and it can be used to identify important aspects of the xc functional for excitonic effects.
It clearly shows that excitons are collective
excitations of the many-body system: the appropriate phase-coherent mixing of the single-particle excitations is
accomplished via a coupling matrix featuring $f_{\rm xc}$. The properties of this coupling matrix are analyzed
and compared with its BSE counterpart.

In textbooks, excitons are usually introduced through the two-body Wannier equation, which describes an electron and a
hole which interact via a screened Coulomb potential. While this arguably constitutes the simplest model
for excitons, it is based on several drastic assumptions which are not fulfilled in general.\cite{S66} We discuss how and under what
circumstances a Wannier-like equation emerges from our minimal TDDFT model.

The paper is structured as follows. We give an introduction to Wannier excitons in Sec. \ref{sect:bkgnd} and to TDDFT
for solids in Sec. \ref{sect:TDDFT}, followed by a description of the minimal model in Sec.
\ref{sect:model}. Section \ref{sect:results} then presents results for the minimal exciton model comparing TDDFT with the BSE,
and discusses various implications. In Sec. \ref{sect:wannier} we show how the Wannier equation emerges from TDDFT.
Conclusions are given in in Sec. \ref{sect:app:details}. Details of the BSE method are provided in Appendix \ref{sect:app:BSE}.
Atomic units ($\hbar=e=m_e=(4\pi\epsilon_0)^{-1}=1$) are used throughout unless otherwise stated,
and we will only consider spin-unpolarized systems.

\section{Background and Model}

\subsection{Wannier excitons} \label{sect:bkgnd}

The electronic structure of crystalline solids is described by the Bloch theory, where the
electrons move in a periodic effective single-particle potential which reflects the crystal symmetry.
As a result, the electronic states form energy bands. In insulators and semiconductors,
electronic excitations take place between the occupied (valence) and unoccupied (conduction) bands.
These interband transitions can be described within a simple independent-particle approach based on Fermi's Golden Rule;
one thus obtains a reasonable qualitative account of the optical properties in these materials.\cite{YC10}

\begin{figure}
\includegraphics[width=0.8\columnwidth]{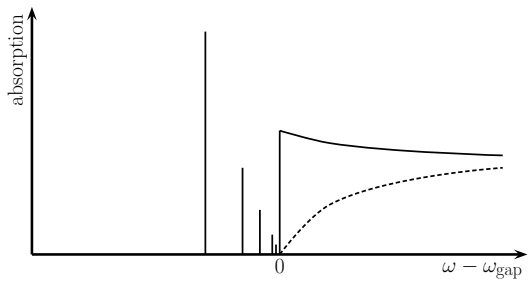}
\caption{Schematic optical spectrum of a typical 3D direct-gap insulator.
Dashed line: independent-particle spectrum. Solid lines:
spectrum including excitonic effects.} \label{fig:bkgnd:IllustrationExcitons}
\end{figure}

However, experiments reveal that there are important modifications to this picture, as illustrated schematically
in Fig. \ref{fig:bkgnd:IllustrationExcitons}. Above the band gap, the spectrum appears strongly enhanced,
and below the band gap one may find discrete absorption peaks known as bound excitons. The origin of these modifications
are Coulomb interactions: the simple picture of independent single-particle excitations is replaced
by a more complex scenario where these excitations are dynamically coupled. Excitonic effects are
ubiquitous in nature, and occur in 3D, 2D and 1D systems alike.\cite{Koch2006,Scholes2006} The details of the
excitonic modifications to the noninteracting spectrum depend on the dimensionality of the system.\cite{HK09}

\begin{figure}
\includegraphics[width=0.55\columnwidth]{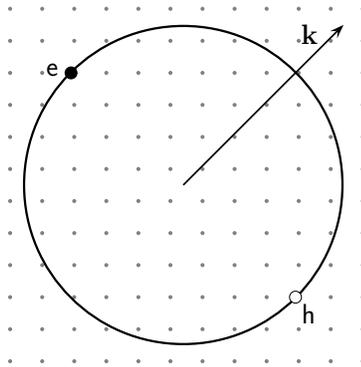}
\caption{Illustration of Wannier exciton as an electron-hole pair, extending over many lattice constants.}
\label{fig:bkgnd:WannierExciton}
\end{figure}

The standard textbook explanation of excitons is based on the simple picture of an electron-hole pair
held together by Coulomb interactions, see Fig. \ref{fig:bkgnd:WannierExciton}. One thus arrives at
a two-body problem similar to the positronium atom, with a center-of-mass momentum $\bf k$ and relative motion described by
the Wannier equation:\cite{HK09}
\begin{equation}
\left[-\frac{\nabla^2}{2m_r}+V(\br)\right]\psi_\nu(\br)=E_\nu\psi_\nu(\br),
\label{eqn:bkgnd:Wannier}
\end{equation}
where $m_r$ is the reduced mass, defined as $m_r^{-1}=m_c^{-1}-m_v^{-1}$.
$m_c$ and $m_v$ are the effective masses of the conduction
band electrons and the valence band holes. $V(\br)=1/\epsilon r$ is the
Coulomb interaction between the electron and the hole, divided by the static dielectric constant of the system.
$\psi_\nu$ and $E_\nu$ are the excitonic wave function and binding energy, respectively.

In 1D systems, the Coulomb interaction is ill-defined and requires, in general, some parametrized form;\cite{GV05}
we will use the following soft-Coulomb interaction:
\begin{equation}
V_\text{1D}(x)=\frac{A}{\sqrt{x^2+\alpha}},
\label{eqn:bkgnd:softcoulomb}
\end{equation}
where $A$ and $\alpha$ are parameters. In the following, we will use $\alpha=0.01$ throughout.

The Wannier equation (\ref{eqn:bkgnd:Wannier}) has the
form of a hydrogenic Schr\"{o}dinger equation, so it possesses a Rydberg series
(even for 1D cases with soft-Coulomb interaction) with infinitely many eigenvalues
below the band gap, and a continuum above the gap.\cite{HK09}
The excitons in the Wannier model for 3D and 2D cases both enhance the optical absorption
near the band gap, while the presence of excitons in 1D systems suppresses the optical absorption
just above the band gap.

Sham and Rice\cite{S66} showed how the Wannier equation can be derived from first principles starting from the BSE,
under the assumption that the Bohr radius of the exciton is much bigger than the lattice constant.
The resulting Wannier picture of excitons appears clear and intuitive, but this simplicity is somewhat deceptive.
In reality, excitons are a dynamical many-body phenomenon and require a subtle coordination and
cooperation of many single-particle transitions between two bands. In the following, we will develop a model
based on TDDFT which will illustrate the true physical nature of excitons, but which will remain sufficiently transparent
to allow a simple interpretation of the collective many-body effects that are responsible for the excitonic binding.

\subsection{TDDFT in finite and periodic systems} \label{sect:TDDFT}

TDDFT\cite{U12} is an in principle exact approach for electron dynamics, based on the uniqueness
of the mapping between the time-dependent electronic density and the external potential.\cite{RG84}
The key equation of TDDFT is the time-dependent Kohn-Sham (TDKS) equation:
\begin{eqnarray}
i\frac{\partial}{\partial t} \phi_i(\br,t)&=&
\left[-\frac{\nabla^2}{2}+v\ext(\br,t)+v_\text{H}(\br,t)\right. \nonumber\\
&&{} +v\xc(\br,t)\bigg]\phi_i(\br,t) ,
\label{eqn:bkgnd:TDDFT}
\end{eqnarray}
where $\phi_i$ are the TDKS orbitals of the noninteracting Kohn-Sham system which reproduces the density
of the real interacting system.  $v\ext$ and $v_\text{H}$
are the external potential of the physical system and the Hartree potential, respectively,
and the xc potential $v\xc$ is the only piece that needs to be approximated in practice.

The excitation spectrum of a system can be calculated via time propagation of Eq. \parref{eqn:bkgnd:TDDFT} following
a suitably chosen initial perturbation.
Alternatively, one can obtain excitation energies and optical spectra directly from linear-response TDDFT,
using the so-called Casida equation:\cite{C96}
\begin{equation}
\left(\begin{array}{cc}\mathbf{A} & \mathbf{B}\\ \mathbf{B} & \mathbf{A}\end{array}\right)
\left(\begin{array}{c}\mathrm{X}\\ \mathrm{Y}\end{array}\right)=
\omega\left(\begin{array}{cc}-\mathbf{1} & \mathbf{0}\\ \mathbf{0} & \mathbf{1}\end{array}\right)
\left(\begin{array}{c}\mathrm{X}\\ \mathrm{Y}\end{array}\right).
\label{eqn:bkgnd:Casida}
\end{equation}
Equation \parref{eqn:bkgnd:Casida} is a generalized eigenvalue equation. One obtains the optical
transition frequencies from the eigenvalues $\omega$, and the corresponding eigenvectors tell
us how the Kohn-Sham single-particle transitions are mixed to form the transitions of the interacting system.
The matrices $\mathbf{A}$ and $\mathbf{B}$ in Eq. \parref{eqn:bkgnd:Casida} are defined as
\begin{equation}
\begin{split}
A^{(ij)(mn)}(\omega)&=\delta_{im}\delta_{jn}(\epsilon_j-\epsilon_i)+F\Hxc^{(ij)(mn)},\\
B^{(ij)(mn)}(\omega)&=F\Hxc^{(ij)(mn)},
\end{split}
\label{eqn:bkgnd:AB}
\end{equation}
where $i$,$j$,$m$,$n$ are labels for ground-state
Kohn-Sham orbitals, and the $\epsilon$'s are the associated Kohn-Sham orbital energies.
$F^{(ij)(mn)}\Hxc$ in Eq. \parref{eqn:bkgnd:AB} is defined as
\begin{multline}
F^{(ij)(mn)}\Hxc=\int\intd^3r\int\intd^3r'\;\phi_i(\vect{r})\phi_j^*(\vect{r})f\Hxc(\vect{r},\vect{r}',\omega)\\
\times\phi_m^*(\vect{r}')\phi_n(\vect{r}'),
\label{eqn:bkgnd:Fxc}
\end{multline}
where the $\phi$'s are the ground-state Kohn-Sham orbitals, and $f\Hxc$ is the Hartree-exchange-correlation (Hxc) kernel,
defined as a Fourier transform of
\begin{equation}
\begin{split}
f\Hxc(\vect{r},t,\vect{r}',t')&=\frac{\delta v_\text{H}(\vect{r},t)}{\delta n(\vect{r}',t')}+\frac{\delta v\xc(\vect{r},t)}{\delta n(\vect{r}',t')}\\
&\equiv \frac{1}{|{\bf r} - {\bf r}'|}+f\xc(\vect{r},t,\vect{r}',t').
\end{split}
\label{eqn:bkgnd:fhxc}
\end{equation}
The xc kernel $f\xc$ has to be approximated in practice.

$\mathrm{X}$ and $\mathrm{Y}$ make up the eigenvector in Eq. \parref{eqn:bkgnd:Casida},
and they describe excitations and de-excitations, respectively. A commonly used approximation
to Eq. \parref{eqn:bkgnd:Casida}, known as Tamm-Dancoff approximation (TDA),\cite{FW03}
is to set $\mathbf{B}=\mathbf{0}$, so that the Casida equation reduces to
\begin{equation}
\sum_{(mn)}\left[\delta_{im}\delta_{jn}(\epsilon_j-\epsilon_i)+F\Hxc^{(ij)(mn)}(\omega)\right]\rho_{mn}(\omega)=\omega\rho_{ij}(\omega).
\label{eqn:bkgnd:TDDFTworking}
\end{equation}
This decouples excitations and de-excitations, and
the computational cost is reduced. There are situations, for instance for molecular excitations of open-shell sys\-tems,\cite{C00}
in which the TDA is preferred in practice over the full Casida equation \parref{eqn:bkgnd:Casida}.
We find that the TDA can also be advantageous for excitons (see Sec. \ref{sect:results}).

By considering only a single Kohn-Sham transition in Eq. \parref{eqn:bkgnd:Casida} one arrives at the small-matrix
approximation (SMA):\cite{GKG97,AGB03}
\begin{equation}
\omega_{\text{SMA},ij}^2=\omega_{\text{KS},ij}^2+4\omega_{\text{KS},ij} F\Hxc^{(ij)(ij)},
\label{eqn:bkgnd:SMA}
\end{equation}
where $\omega_{\text{KS},ij}=\epsilon_j-\epsilon_i$ is the Kohn-Sham transition frequency to be corrected. One can
further simplify this by making the TDA, which yields the single-pole approximation (SPA):
\begin{equation}
\omega_{\text{SPA},ij}=\omega_{\text{KS},ij}+ 2F\Hxc^{(ij)(ij)}.
\end{equation}
The SMA and SPA are valid when the considered excitation is far away from other transitions in the system.
Though they are not usually accurate enough for real calculations, their simplicity makes them very useful for theoretical
analysis and development.

In periodic solids, the Kohn-Sham orbitals are labeled with the band index $i$ and the wavevector $\vect{k}$ and have the Bloch form:
\begin{equation}
\phi_{i\vect{k}}(\vect{r})=e^{i\vect{k}\cdot\vect{r}}u_{i\vect{k}}(\vect{r}).
\label{eqn:bkgnd:BlochForm}
\end{equation}
It is in principle possible to adapt the Casida equation \parref{eqn:bkgnd:Casida} for the case of orbitals of the
form (\ref{eqn:bkgnd:BlochForm}),\cite{Gruning2007,Izmaylov2008} and use this to calculate the excitation spectrum. However,
to obtain the optical spectrum in solids it is more convenient to calculate the macroscopic dielectric function:\cite{ORR02}
\begin{equation}
\begin{split}
\epsilon_M(\omega)&=\lim_{q\to0}\frac{1}{\epsilon^{-1}_{G=G'=0}(\vect{q},\omega)}\\
&=\lim_{q\to0}\frac{1}{1+v_{G=0}(\vect{q})\chi_{G=G'=0}(\vect{q},\omega)},
\end{split}
\label{eqn:bkgnd:macrodielec}
\end{equation}
where the $\vect{G}$'s are reciprocal lattice vectors, and $\chi=\delta n/\delta v_\text{ext}$
is the linear response function. $\epsilon_M(\omega)$ can be expressed as\cite{ORR02}
\begin{equation}
\epsilon_M(\omega)=1-\lim_{\vect{q}\to0}v_{G=0}(\vect{q})\sum_{\lambda}
\frac{\abs{\sum_{ij}\matelem{i}{e^{-i\vect{q}\cdot\vect{r}}}{j}}^2}{\omega-\omega_\lambda+i\eta},
\label{eqn:bkgnd:macrodielecworking}
\end{equation}
where $\lambda$ labels the solutions of Eq. \parref{eqn:bkgnd:TDDFTworking} for an extended system.

Beyond linear response, few real-time TDDFT calculations exist for periodic solids.\cite{Bertsch2000,Otobe2008}
Instead of directly solving the TDKS equation (\ref{eqn:bkgnd:TDDFT}),
the TDKS orbitals can be expanded in terms of ground-state Kohn-Sham Bloch functions as
\begin{equation}
\phi_{i\vect{k}}(\vect{r},t)=\sum_m c_{im\vect{k}}(t)\phi_{m\vect{k}}(\vect{r}).
\end{equation}
The time-dependent density matrix is defined as
\begin{equation} \label{tddm}
\rho_{i\vect{k}}^{mn}(t) =c_{im\vect{k}}(t)c_{in\vect{k}}^*(t).
\end{equation}
The equation of motion of the density matrix is then
\begin{equation}
i\frac{\partial}{\partial t}\bm{\rho}_{i\vect{k}}(t)=\left[\vect{H}_\vect{k}(t),\bm{\rho}_{j\vect{k}}(t)\right],
\label{eqn:bkgnd:TDDMEOM}
\end{equation}
with the TDKS Hamiltonian matrix $\vect{H}_\vect{k}(t)$ defined by
\begin{equation} \label{hmat}
{H}_\vect{k}^{mn}(t)=\int_\Omega\intd^3r\;\phi^*_{m\vect{k}}(\vect{r})H_{\rm KS}(\vect{r},t)\phi_{n\vect{k}}(\vect{r}),
\end{equation}
where $\Omega$ is the volume of the unit cell, and $H_{\rm KS}(\vect{r},t)$ is the TDKS Hamiltonian of Eq. (\ref{eqn:bkgnd:TDDFT}).
This density-matrix approach has been used to derive the
TDDFT version of the semiconductor Bloch equations.\cite{TU08,TLU09}
In this formalism we only consider vertical transitions, where the Bloch wavevector $\bf k$
does not change during the dynamics. Nonvertical excitations are not considered, since they involve indirect (i.e., phonon-assisted)
processes which we ignore here.

\subsection{Minimal model for excitons}
\label{sect:model}

Solids are formally described by the many-body Schr\"{o}\-dinger equation. Since exact solutions are not possible,
it is instructive to resort to model systems to eliminate undesired details of the many-body system and
provide clear illustrations of the specific features one is interested in. The Wannier equation for excitons presented in
Sec. \ref{sect:bkgnd} is such a model.

While intuitive, the Wannier model assumes the elec\-tron-hole interaction as given,
so the excitonic effects are already built in by default.
However, this does not explain under what conditions one
expects to see the formation of excitons, and the many-body nature of excitonic effects remains hidden.
We therefore propose a minimal model for excitons which
lowers the abstraction level of the Wannier model, and where excitonic effects show up without any {\em ad hoc} assumptions.

For excitations near the band gap, a reasonable approximation is to use a two-band model, i.e., only to consider
the highest valence band $(v)$ and the lowest conduction band $(c)$. This means that we only need to consider those
elements of the time-dependent density matrix $\rho_{i\vect{k}}^{mn}(t)$ [Eq. (\ref{tddm})] for which  $mn = cc, cv, vc, vv$
and $i=v$; the latter index will be dropped in the following.

For the case when a small perturbative electric field is applied to the system, it is sufficient, to lowest order in the perturbation,
to consider only the time evolution of the off-diagonal part of the density matrix,\cite{TLU09} $\rho_{\vect{k}}^{cv}$.
One then obtains from Eq. \parref{eqn:bkgnd:TDDMEOM}
\begin{equation}
i\frac{\partial}{\partial t}\rho_\vect{k}^{cv}(t)=\omega_\vect{k}^{cv}\rho_\vect{k}^{cv}(t)+\delta V_{{\sss HXC},\vect{k}}^{cv}(t),
\label{eqn:bkgnd:rhocvInTime}
\end{equation}
where $\omega_\vect{k}^{cv}=\epsilon_{c\vect{k}}-\epsilon_{v\vect{k}}$, and
$\delta V_{{\sss HXC}, \vect{k}}^{cv}(t)=V_{{\sss HXC}, \vect{k}}^{cv}(t)-V_{{\sss HXC}, \vect{k}}^{cv}(0)$.
Here, $V_{{\sss HXC},\vect{k}}^{cv}$ denotes the matrix elements of $v_\text{H}(\br,t) +v\xc(\br,t)$, defined similarly to
Eq. (\ref{hmat}).
There is no external perturbation in Eq. \parref{eqn:bkgnd:rhocvInTime}; we assume free propagation of the
system. Our reasoning is that individual excitations can be viewed as the eigenmodes of the system and do not
depend on the specific form of the perturbative field.
Fourier transformation of Eq. \parref{eqn:bkgnd:rhocvInTime} gives
\begin{equation}
\rho_\vect{k}^{cv}(\omega)=\frac{2\sum_{\vect{k}'}\left\{F_{{\sss HXC},\vect{k},\vect{k}'}^{(vc)(cv)}\rho_{\vect{k}'}^{cv*}(\omega)
+F_{{\sss HXC},\vect{k},\vect{k}'}^{(vc)(vc)}\rho_{\vect{k}'}^{cv}\right\}}{\omega-\omega_\vect{k}^{cv}},
\label{eqn:bkgnd:rhocv}
\end{equation}
where the factor 2 accounts for the spin, and
\begin{equation}
\begin{split}
F_{{\sss HXC}, {\vect{k},\vect{k}'}}^{(ij)(mn)}&=\int_{\Omega}\intd^3r\int_{\Omega}\intd^3r'\;\phi_{i\vect{k}}(\vect{r})\phi_{j\vect{k}}^*(\vect{r})\\
&\quad\quad\times f\Hxc(\vect{r},\vect{r}',\omega)\phi_{m\vect{k}'}^*(\vect{r}')\phi_{n\vect{k}'}(\vect{r}')\\
&\equiv \langle \langle ij|f\Hxc|mn \rangle \rangle.
\end{split}
\label{eqn:bkgnd:Fijmn}
\end{equation}

Equation \parref{eqn:bkgnd:rhocv} is equivalent to the SMA for finite systems. While the SMA only refers to
the transition between one individual occupied and one individual unoccupied orbital, Eq. \parref{eqn:bkgnd:rhocv}
considers the transitions between the valence and the conduction band as a whole.
Ignoring the coupling between excitations and de-excitations by setting $F_{\vect{k},\vect{k}'}^{(vc)(cv)}=0$ (i.e., making the TDA),
one arrives at the solid-state analog of the SPA:
\begin{equation}
\sum_{\vect{k}'}\left[\omega_{\vect{k}'}^{cv}\delta_{\vect{k},\vect{k}'}+F_{{\sss HXC},
\vect{k},\vect{k}'}^{(vc)(vc)}(\omega)\right]\rho_{\vect{k}'}^{cv}(\omega)
=\omega\rho_\vect{k}^{cv}(\omega).
\label{eqn:bkgnd:rhocvSPA}
\end{equation}
This is the central equation which we will use to describe excitonic effects. It requires as input the Kohn-Sham
Bloch functions for the valence and the conduction band of an insulator or semiconductor. To keep things as
simple as possible, we will consider the Kronig-Penney (KP) model\cite{K05} rather than a real material.
The KP model is a 1D noninteracting system with a periodic potential of square wells.
Within the unit cell $[-b,a]$, the potential is
\begin{equation}
V_\text{KP}(x)=\left\{\begin{array}{cc}0 & 0<x<a\\ V_0 & -b<x<0 \:.\end{array}\right.
\label{eqn:bkgnd:KPpot}
\end{equation}

\begin{figure}
\includegraphics[width=0.85\columnwidth]{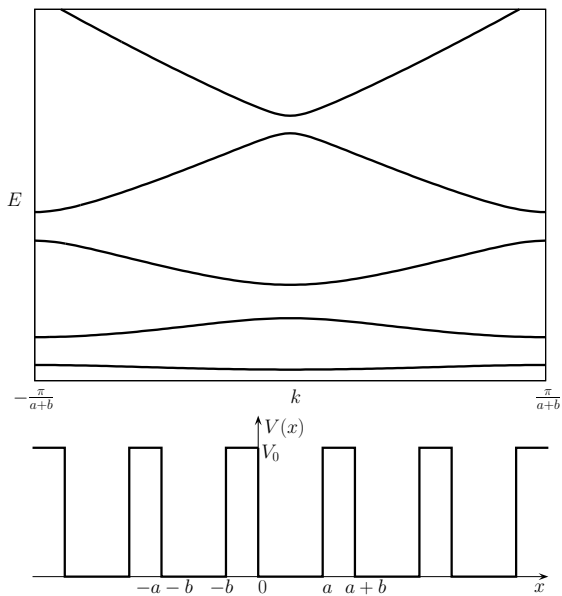}
\caption{1D Kronig-Penney model: potential (lower panel) and band structure (upper panel).}
\label{fig:bkgnd:KPpot}
\end{figure}

A typical example for the band structure of the KP model is plotted in Fig. \ref{fig:bkgnd:KPpot}.
Despite its simple appearance, the KP model is very versatile: by varying the values of the lattice constant
$a+b$, barrier width $b$, and barrier height $V_0$, a wide range of band gaps and band
curvatures can be achieved. A square-well potential of finite depth does not support an infinite number of bound
states as the Coulomb potential does; but this, in fact, closely reflects the reality of the effective potential
felt by the valence electrons in a solid, which is relatively shallow due to the screening of the bare nuclear charges by the core
electrons. In practice, many solid-state calculations account for this screening by using muffin-tin potentials or
pseudopotentials. \cite{PK59,KP60,FNM03} The KP model can be viewed as an elementary version of this approach.

In the following, we always choose the first two bands to be fully occupied and the higher bands to be empty. We then make the
two-band approximation for band 2 and 3, since the shapes of these bands resemble the highest
valence band and lowest conduction band in direct-gap materials such as GaAs. The bands in the KP model
are sufficiently well separated, so the two-band approximation is justified.

To establish a connection with TDDFT, we assume the solution to the noninteracting KP model as our ground-state
Kohn-Sham system. In other words, the potential in Eq. \parref{eqn:bkgnd:KPpot} represents the {\em exact} Kohn-Sham potential
$v_{\rm ext} + v_{\rm H} + v_{\rm xc}$ which corresponds to a physical system whose external potential $v_{\rm ext}$ is
uniquely determined thanks to the Hohenberg-Kohn theorem of DFT. For our purpose, it is not necessary to know what this
external potential looks like. The Kohn-Sham Bloch
functions can then be determined in an elementary fashion.\cite{K05}

For comparison, we will also carry out BSE calculations in our model system (see Appendix \ref{sect:app:BSE} for technical details).
BSE calculations are typically based on ground-state quasiparticle states obtained from
the GW method.\cite{ORR02} This is because the single-particle gap in GW is usually closer to experiment than the approximate Kohn-Sham gap.
However, in our case this distinction is not important because we use the given KP band structure as input for both BSE and TDDFT.

In summary, our minimal TDDFT model for excitons consists of the following two ingredients:

(1) A two-band model for the vertical transitions between the highest valence band and the lowest conduction band, see
Eq. (\ref{eqn:bkgnd:rhocvSPA});

(2) the band structure from a 1D KP model.

Of course, the model is not complete without a choice for the xc kernel $f\xc$. This will be discussed below.

\section{Results from the minimal model}
\label{sect:results}

\subsection{Bound excitons from the BSE and from TDDFT}

The exact xc kernel $f\xc(\vect{r},\vect{r}',\omega)$ is unknown and must be approximated;
we restrict ourselves to adiabatic kernels that have no frequency dependence.
The adiabatic local-density approximation (ALDA), as well as all semilocal, gradient-corrected xc kernels,
are known to be unable to describe excitonic effects.\cite{ORR02} The exact xc kernel
has a long-range decay of $1/\abs{\vect{r}-\vect{r}'}$, which is absent in all (semi)local xc kernels
derived from the uniform electron gas. This long-range part is
thought to be essential for excitons.\cite{ORR02,U12,DAOR03}

The long-range behavior of $f\xc$ depends on the dimensionality, and in our 1D model system we define the
following long-ranged (or `soft-Coulomb') xc kernel:
\begin{equation}
f\xc^\text{SC}(x,x',\omega)=-\frac{A^\text{SC}}{\sqrt{(x-x')^2+\alpha}}.
\label{eqn:results:fxcsc}
\end{equation}
We also consider an extremely short-ranged contact xc kernel:
\begin{equation}
f\xc^\text{cont}(x,x',\omega)=-A^\text{cont}\delta(x-x').
\label{eqn:results:fxccont}
\end{equation}
These model xc kernels depend on the constants $A^\text{SC}$ and $A^\text{cont}$,
which we will treat as fitting parameters in the following. The idea is to tune the parameters in the
model xc kernels so that bound excitons are produced, and to align the lowest bound exciton in the TDDFT spectrum with
the lowest bound exciton in the BSE spectrum.

\begin{figure}
\includegraphics[height=\columnwidth,angle=-90]{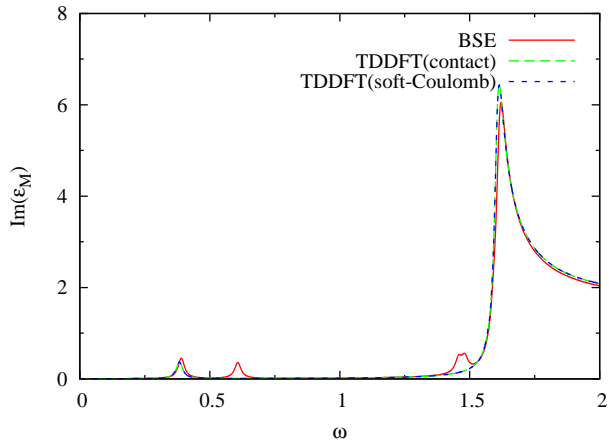}
\caption{Imaginary part of the dielectric function, $\mathrm{Im}(\epsilon_M)$, calculated with BSE and TDDFT.
Parameters of the KP model: $a=2.6$, $b=0.4$, $V_0=8$.
For BSE, $A=0.25$ in Eq. \parref{eqn:bkgnd:softcoulomb}.
For TDDFT with the contact kernel, Eq. \parref{eqn:results:fxccont}, $A^\text{cont}=2.32$.
For TDDFT with the soft-Coulomb
kernel, Eq. \parref{eqn:results:fxcsc}, $A^\text{SC}=0.898$. The BSE produces several bound
excitons, but TDDFT only one.}
\label{fig:results:TDDFTOneExciton}
\end{figure}

\begin{figure}
\includegraphics[height=\columnwidth,angle=-90]{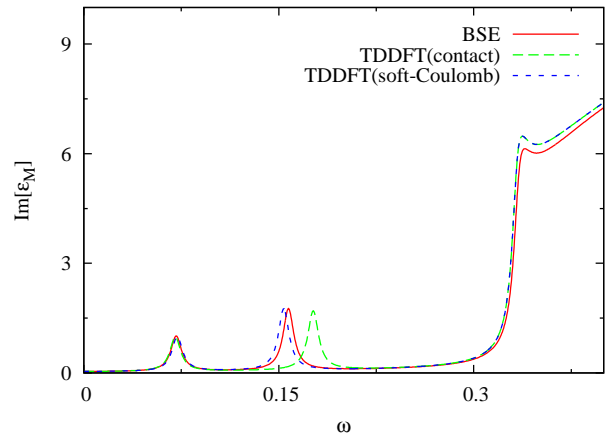}
\caption{Same as Fig. \ref{fig:results:TDDFTOneExciton}, for KP model parameters $a=3$, $b=3$, $V_0=1$.
For BSE, $A=0.14$ in Eq. \parref{eqn:bkgnd:softcoulomb}.
For TDDFT with the contact kernel, Eq. \parref{eqn:results:fxccont}, $A^\text{cont}=3.77$.
For TDDFT with the soft-Coulomb
kernel, Eq. \parref{eqn:results:fxcsc}, $A^\text{SC}=0.955$. TDDFT produces
two bound excitons. Higher-lying bound excitons exist within BSE but are numerically hard to resolve.}
\label{fig:results:TDDFTTwoExcitons}
\end{figure}

Results for the imaginary part of the dielectric function are presented in Figs. \ref{fig:results:TDDFTOneExciton}
and \ref{fig:results:TDDFTTwoExcitons}. We find that both the long-ranged $f\xc^\text{SC}$ and the short-ranged
$f\xc^\text{cont}$ produce bound excitons, and thus, strictly speaking, the long-range behavior of the xc kernel
is not really required for excitonic effects. The BSE results in
Figs. \ref{fig:results:TDDFTOneExciton} and \ref{fig:results:TDDFTTwoExcitons}
show several identifiable bound excitons, in agreement with the Rydberg series
predicted by the Wannier model; the number of visible bound excitons somewhat depends on the numerical resolution in momentum space.

For the KP model parameters of Fig. \ref{fig:results:TDDFTOneExciton}, we find that the adiabatic TDDFT
can only bind a single excitonic state. For other KP parameters (specifically those in which
the lowest conduction band is above the barrier),
TDDFT produces two excitons, see Fig. \ref{fig:results:TDDFTTwoExcitons}, which agree well with the lowest two excitons in
BSE. There are additional, higher-lying bound excitons in BSE which are very faint and difficult to resolve numerically.
For all the KP systems we tested, we never found more than two bound excitons with TDDFT.
This indicates the limitations of the adiabatic xc kernels used here.

\begin{figure}
\includegraphics[height=\columnwidth,angle=-90]{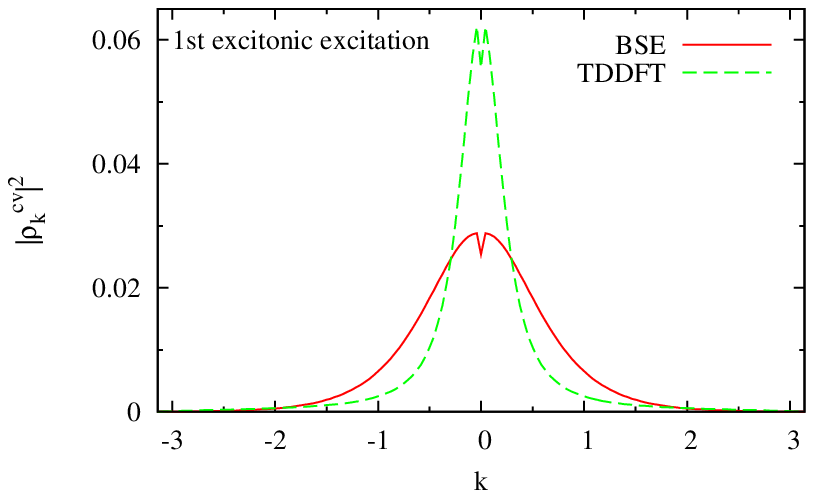}\\
\includegraphics[height=\columnwidth,angle=-90]{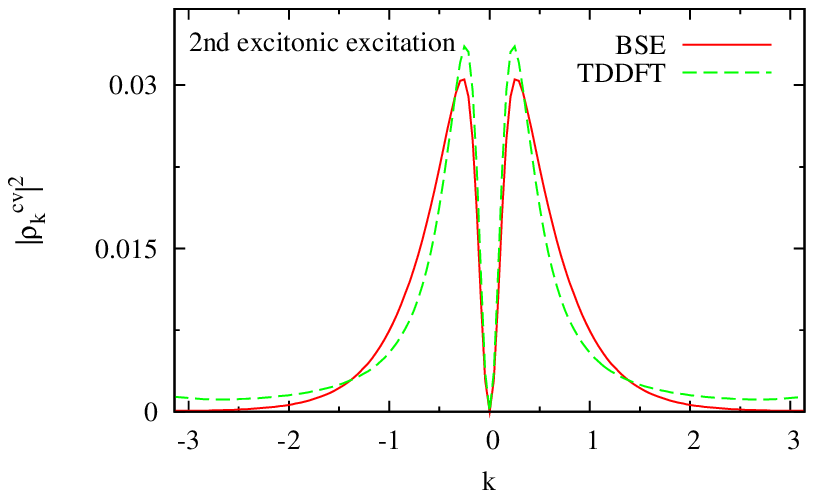}
\caption{Eigenvectors $\abs{\rho_k^{cv}}^2$ of the first two excitonic transitions for KP
parameters $a=0.5$, $b=0.5$, $V_0=20$. For BSE,
$A=0.25$ in Eq. \parref{eqn:bkgnd:softcoulomb}.
For TDDFT (soft-Coulomb), $A^\text{SC}=2.39$ in Eq. \parref{eqn:results:fxcsc}.}
\label{fig:results:ExcitonicEigenvect}
\end{figure}

\begin{figure}
\includegraphics[height=\columnwidth,angle=-90]{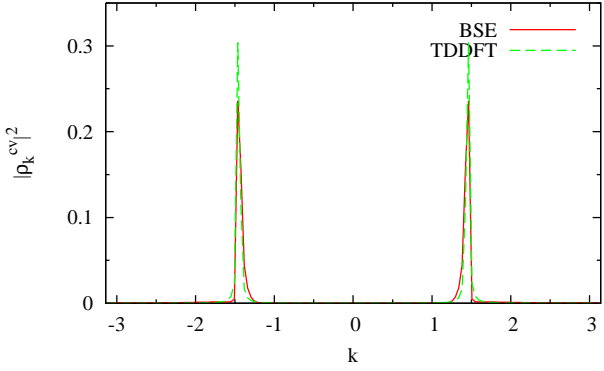}
\caption{Eigenvector $\abs{\rho_k^{cv}}^2$ of a nonexcitonic excitation in the continuum
part of the spectrum. The model parameters are the same
as those in Fig. \ref{fig:results:ExcitonicEigenvect}.}
\label{fig:results:NonExcitonicEigenvect}
\end{figure}

As mentioned in Sec. \ref{sect:model}, the Wannier model does not clearly demonstrate
that excitons are collective excitations. Since the Wannier model assumes a single
electron-hole pair picture, one cannot immediately see that excitonic excitations
are composed of a coherent superposition of many single-particle excitations. In our minimal model,
we solve the eigenvalue equation \parref{eqn:bkgnd:rhocvSPA}, and the eigenvectors
$\rho_k^{cv}$ (which depends on $\omega$ parametrically) describe how the transitions
between noninteracting orbitals form the transitions in the interacting system.
$\abs{\rho_k^{cv}}^2$ is the percentage of a noninteracting transition in the transition
of the interacting system. Two typical cases are plotted in
Figs. \ref{fig:results:ExcitonicEigenvect} and \ref{fig:results:NonExcitonicEigenvect}.

Figure \ref{fig:results:ExcitonicEigenvect} clearly shows that excitons are
collective excitations which are formed by mixing a wide distribution of single-particle transitions.
As expected, the lowest exciton eigenfunction is nodeless and the second excitonic eigenfunction
has a single node. With purely parabolic bands,
the results from the Wannier model would be recovered, as we will show below. In contrast, the transitions
in the continuum  shown in Fig. \ref{fig:results:NonExcitonicEigenvect} have a strong single-particle character
(the two peaks arise from the $\pm k$ degeneracy in the KP model).

Equation \parref{eqn:bkgnd:rhocvSPA} is equivalent to the SPA for finite systems, which
ignores the coupling between excitations and de-excitations (TDA). We also investigated
what happens when we do not make the TDA, i.e., when we work instead of Eq. \parref{eqn:bkgnd:rhocvSPA}
with the full equation for the two-band model, Eq. \parref{eqn:bkgnd:rhocv}. As long as we describe relatively weakly
bound excitons that are not too far below the band gap, we find that the difference between the two methods is very minor.

However, we also discovered that, under rare circumstances,  Eq. \parref{eqn:bkgnd:rhocv} can lead to
TDDFT excitonic binding energies that are purely imaginary.
In our minimal model, such instabilities arise when the interaction strength
$A$ in Eq. \parref{eqn:results:fxcsc} and \parref{eqn:results:fxccont} increases so that the excitonic binding energy becomes greater than the band gap. This situation is comparable to the well-known
triplet instability in TDDFT, for which the TDA generally leads to an overall better behavior;\cite{C00}
for excitons binding energies in our minimal model, we draw similar conclusions.

\subsection{Analysis of the coupling matrix}

\begin{figure}
\includegraphics[height=\columnwidth,angle=-90]{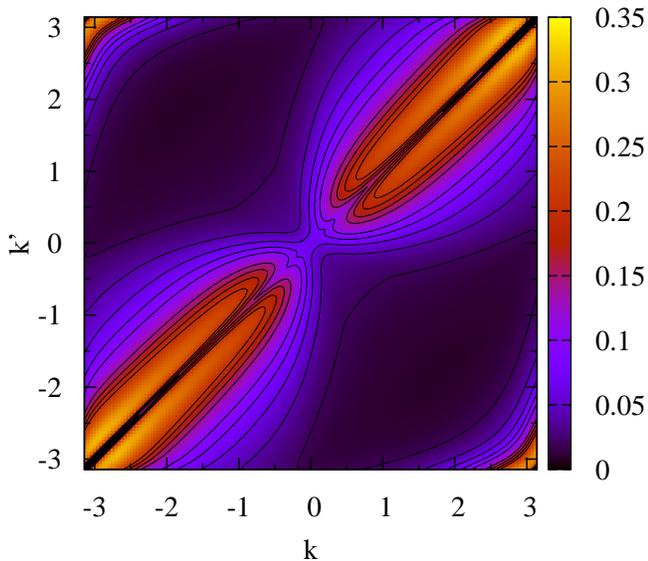}
\caption{Contour plot of the coupling matrix $\abs{F_\text{BSE}^{(vc)(vc)}}$. The model parameters
are the same as those in Fig. \ref{fig:results:ExcitonicEigenvect}.}
\label{fig:results:BSEcoupling}
\end{figure}

\begin{figure}
\includegraphics[height=\columnwidth,angle=-90]{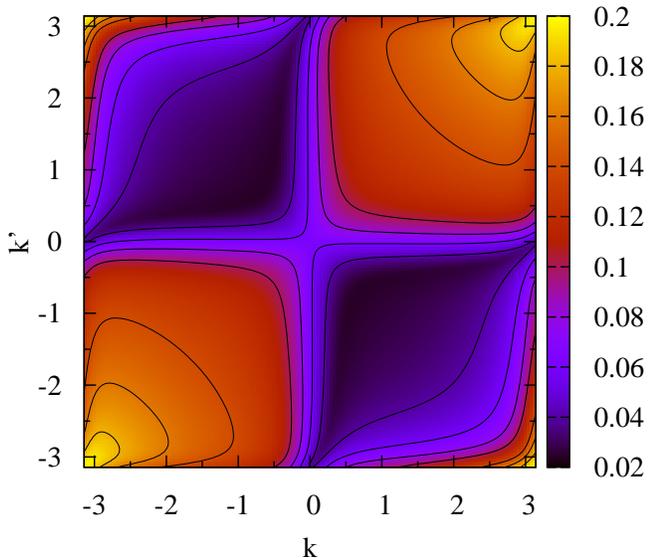}
\caption{Contour plot of the coupling matrix $\abs{F_\text{TDDFT}^{(vc)(vc)}}$. The model parameters
are the same as those in Fig. \ref{fig:results:ExcitonicEigenvect}.}
\label{fig:results:TDDFTcoupling}
\end{figure}

The BSE scheme is commonly implemented within an adiabatic scheme (see Appendix \ref{sect:app:BSE}); as we have seen,
it produces a series of bound excitons. Since we assumed that the KP model is the Kohn-Sham ground state in TDDFT and
the GW quasiparticle ground state in BSE, the difference between TDDFT and BSE becomes easily comparable,
since the central equation to be solved have the same form, Eq. \parref{eqn:bkgnd:rhocvSPA}.
The $F^{(ij)(mn)}$ coupling matrices for TDDFT and BSE are
\begin{align}
\!\!\!\!F_{\text{TDDFT},\vect{k},\vect{k}'}^{(ij)(mn)}&=2\langle \langle ij | f_\text{H} | mn \rangle \rangle +
2\langle \langle ij | f\xc | mn \rangle \rangle,\\
F_{\text{BSE},\vect{k},\vect{k}'}^{(ij)(mn)}&=2\langle \langle ij | f_\text{H} | mn \rangle \rangle -
\langle \langle im | W | jn \rangle \rangle,
\label{eqn:results:FBSE}
\end{align}
where $f_\text{H}$ is the Hartree kernel (the 1D soft-Coulomb interaction),
and $W$ is the screened interaction. Aside from the
change from $f\xc$ to $W$, the most prominent difference between BSE and TDDFT is the order of the indices
for $W$ in Eq. \parref{eqn:results:FBSE}. Since the noninteracting ground-state wave functions have
the Bloch form \parref{eqn:bkgnd:BlochForm}, we can see from Eq. \parref{eqn:bkgnd:Fijmn}
that $F^{(ij)(mn)}_\text{BSE}$ has a strong $\vect{k}-\vect{k}'$ dependence;
in Fig. \ref{fig:results:BSEcoupling} this shows up as a dominance along the diagonal.
By contrast, this $\vect{k}-\vect{k}'$ dependence is clearly absent in $F^{(ij)(mn)}_\text{TDDFT}$, as demonstrated
in Fig. \ref{fig:results:TDDFTcoupling}.

$\langle\langle ij|f\xc|mn \rangle\rangle$ and $\langle\langle im|W|jn \rangle\rangle$
with only vertical transitions can be expressed in momentum space as
\begin{multline}
\langle\langle ij|f\xc|mn \rangle\rangle=\frac{1}{\Omega}\sum_{\vect{G},\vect{G}'}f\xc(q=0,\vect{G},\vect{G}')\\
\times\matelem{j,\vect{k}}{e^{i\vect{G}\cdot\vect{r}}}{i,\vect{k}}\matelem{m,\vect{k}'}{e^{-i\vect{G}'\cdot\vect{r}}}{n,\vect{k}'},
\label{eqn:results:fxctrans}
\end{multline}
\begin{multline}
\langle\langle im|W|jn \rangle\rangle=\frac{1}{\Omega}\sum_{\vect{G},\vect{G}'}W(\vect{q}=\vect{k}-\vect{k}'+\vect{G}_0,\vect{G},\vect{G}')\\
\times\matelem{j,\vect{k}}{e^{i(\vect{q}+\vect{G})\cdot\vect{r}}}{n,\vect{k}'}
\matelem{m,\vect{k}'}{e^{-i(\vect{q}+\vect{G}')\cdot\vect{r}}}{i,\vect{k}}.
\label{eqn:results:Wtrans}
\end{multline}
The xc matrix \parref{eqn:results:fxctrans} only depends on the long-range ($q=0$)
behavior of its momentum space representation $f\xc(\vect{q},\vect{G},\vect{G}')$,
while the $W$ matrix \parref{eqn:results:Wtrans} also depends on
other $\vect{q}$ values in its momentum space representation $W(\vect{q},\vect{G},\vect{G}')$.
It is impossible to find an
adiabatic $f\xc$ that reproduces the BSE coupling matrix as in Fig. \ref{fig:results:BSEcoupling},
since $W(\vect{q},\vect{G},\vect{G}')$ has an extra degree of freedom over $f\xc(q=0,\vect{G},\vect{G}')$.
One can only hope to reproduce a portion of the BSE coupling matrix with adiabatic
TDDFT (as pointed out in Ref. \onlinecite{RORO02}), or make the xc kernel frequency dependent
so that the information from the $\vect{q}$-dependence in $W(\vect{q},\vect{G},\vect{G}')$
is mapped into the frequency dependence in $f\xc(q=0,\vect{G},\vect{G}',\omega)$.

Considering the nature of the objects involved in this mapping, a highly nontrivial
frequency dependence in $f\xc$ is required to reproduce a series of bound excitons.
For example, one can easily construct an $f\xc$  which
reproduces a given series of bound excitons by using a different contact kernel
in the region $\omega_i-\eta_i^-\le \omega\le \omega_i+\eta_i^+$ surrounding each exciton at frequency $\omega_i$:
\begin{equation}
f\xc(x,x',\omega)= A_i\delta(x-x')\theta[\omega-(\omega_i-\eta_i^-)]\theta[(\omega_i+\eta_i^+)-\omega].
\end{equation}
Here, the $A_i$'s are parameters which are adjusted so that a TDDFT calculation with the frequency-independent
kernel $f\xc^{i}(x,x')= A_i\delta(x-x')$ would produce $\omega_i$ as the lowest excitonic binding energy.
Such an $f\xc$ is of course completely ad hoc, but the fact that the excitonic series can be reproduced in this way
demonstrates that the inclusion of the frequency dependence would
greatly improve the flexibility of the TDDFT scheme.

On the other hand, within the adiabatic approximation the characteristics of
the $F\Hxc^{(ij)(mn)}$ coupling matrix are important for excitonic effects. To
emphasize this, we now show that in very special cases the number of discrete excitonic
eigenvalues can be derived. Consider a real matrix
$\Omega^{(0)}+F$, where $\Omega^{0}$ is a diagonal real matrix with
$\Omega^{0}_{k,k}=\Omega^{(0)}_{-k,-k}=\omega^{(0)}_k$, and $F$ has the symmetry
$F_{k,k'}=F_{-k',-k}=F_{k',k}=F_{-k,k'}=F_{k,-k'}$. Within second order perturbation theory,
there is at most one discrete eigenvalue of $\Omega^{(0)}+F$ in the limit where $k, k'$ become continuous.\cite{supplmat}
Though this case does not correspond to the matrices that would occur in real
calculations, it indicates the close relationship between the properties of the
coupling matrix and excitonic effects.

It is also possible to derive properties of the discrete eigenvalues if $k$ and $k'$
are completely decoupled in the xc kernel. Owing to the symmetry
$F^{(vc)(vc)}_{k,k'}=F^{(vc)(vc)}_{-k',-k}=F^{(vc)(vc)*}_{k',k}$ implied
by Eq. \parref{eqn:bkgnd:Fijmn}, such separable kernels can only have the form
\begin{equation}
F_{k,k'}=\pm A(k)A^*(k').
\label{eqn:results:sepkernel}
\end{equation}
For an excitation below the band gap with frequency $\omega$, we can show\cite{supplmat} that it must satisfy
\begin{equation}
-\sum_k\frac{\abs{A(k)}^2}{\omega-\omega^{cv}_k}\quad=\quad 1,
\label{eqn:results:eqnsepkernel}
\end{equation}
where the sum is carried out over the first Brillouin zone (FBZ). Equation \parref{eqn:results:eqnsepkernel}
shows that Eq. \parref{eqn:results:sepkernel} must have the negative sign in order to have
bound excitons. The left-hand side of Eq. \parref{eqn:results:eqnsepkernel} is
monotonically increasing with $\omega$, so for separable kernels of the form of Eq. \parref{eqn:results:sepkernel},
there is only one bound excitonic solution.

As shown in Fig. \ref{fig:results:TDDFTcoupling},
TDDFT coupling matrices lack the strong dependence of $k-k'$ as in BSE coupling matrices.
Expanding the TDDFT coupling matrices into a power series of separable matrices and
truncating at the first order would be a reasonable approximation, explaining why TDDFT
produces fewer bound excitons (if any at all) than many-body methods such as BSE.

\subsection{Dimensionality considerations}

The contact xc kernel and the soft-Coulomb xc kernel in Eqs. \parref {eqn:results:fxcsc}
and \parref{eqn:results:fxccont} have the following simple form in momentum space:
\begin{equation}
\begin{split}
f\xc^\text{SC}(q,G,G')&=-2A^\text{SC}K_0\left(\sqrt{\alpha^\text{SC}}\abs{q+G}\right)\delta_{G,G'},\\
f\xc^\text{cont}(q,G,G')&=-A^\text{cont}\delta_{G,G'},
\end{split}
\end{equation}
where $q\in\text{FBZ}$, $G$ and $G$' are reciprocal lattice vectors, and $K$ is the
modified Bessel function of the second kind. It is customary to refer to  the matrix elements
where $G=G'=0$ as `head', $G=0$ or $G'=0$ as `wings', and $G\ne0, G'\ne0$ as `body'.

The 3D Coulomb potential has the form $4\pi/q^2$ in momentum space. However, in 1D systems
there is no real Coulomb interaction which behaves as $q^{-2}$ for $q\to0$, and one has
to use the soft-Coulomb interaction instead. Though there are many flavors of the soft-Coulomb
interaction, they all have the same $\log q$ behavior for $q\to0$. However, the linear response
function $\chi$ always behaves as $q^{-2}$ for $q\to0$ and does not depend on the dimensionality.
This renders quantities like the macroscopic dielectric function \parref{eqn:bkgnd:macrodielec}
ill-defined for strictly 1D systems. Furthermore, the bootstrap xc kernel\cite{SDSG11}
and other xc kernels that depend on the cancellation of the 3D Coulomb $q^{-2}$ singularity
will not work as designed in strictly 1D and 2D systems. Therefore $\mathrm{Im}(\epsilon_M)$
shown in Fig. \ref{fig:results:TDDFTOneExciton} and \ref{fig:results:TDDFTTwoExcitons} are calculated at a small but finite $q$.

The coupling matrix $\mathbf{F}_{{\sss HXC}}^{(vc)(vc)}$ can be written in momentum space as
\begin{eqnarray}
F_{{\sss HXC}, k,k'}^{(vc)(vc)}&=&
\frac{1}{\Omega}\sum_{G,G'}\left[v_G(q=0)\delta_{G,G'}+f\xc(q=0,G,G')\right] \nonumber\\
&& \times\matelem{c,k}{e^{iGx}}{v,k}\matelem{v,k'}{e^{-iG'x}}{c,k'}.
\label{eqn:app:detail:Fvcvc}
\end{eqnarray}

For any xc kernel that behaves as $q^{-2}$ for $q\to0$, one can further simplify the calculation
by ignoring the so-called local field effects,\cite{SW02} i.e. instead of summing over $G$ and $G'$ in
Eq. \parref{eqn:app:detail:Fvcvc}, only the head is considered. In 3D systems,
a prominent example is the long-range kernel $-\alpha/q^2$, which is obtained as an effective
xc kernel with only head matrix elements from inverting the BSE of contact excitons.\cite{SKRA03}

On the other hand, any xc kernel
that diverges more slowly than $q^{-2}$ for $q\to0$ changes the spectrum only through the local field effects,
i.e. all $G$ and $G'$ must be summed in Eq. \parref{eqn:app:detail:Fvcvc}. In other words,
effective xc kernels with only the head are not feasible in strictly 1D systems
due to the asymptotic behavior of the soft-Coulomb potential discussed above. For 1D systems with $G=0$, we have
\begin{equation}
\begin{split}
\matelem{j,k_j}{e^{i(q+G)x}}{i,k_i}&\stackrel{q\to0}{\sim}O(q^1),\\
v_{G=0}(q)&\stackrel{q\to0}{\sim}O(\log q),
\end{split}
\label{eqn:app:detail:asymp}
\end{equation}
and $f\xc$'s with 1D long-range behavior such as the soft-Coulomb kernel also behave as $O(\log q)$.
Considering Eq. \parref{eqn:app:detail:Fvcvc}, these asymptotic properties imply that the head and wing
contributions to $F^{(ij)(mn)}$ always vanish in strictly 1D systems for physically meaningful xc kernels. Due to these dimensionality
restrictions, the xc kernel changes the strictly 1D system only through the local field effects.

In 3D the head contribution to the coupling matrix $\mathrm{F}\Hxc$ is much more important
than the local field effects, which is the reason that long-range kernels (with nonzero head)
work much better than local xc kernels (with vanishing head) such as ALDA. In our strictly 1D model system, the head
contribution is zero even for the BSE, and thus the long-range kernel does not outperform local
kernels such as the contact kernel.

These peculiarities only occur when one considers strictly 1D and 2D systems.
In a more realistic picture, one encounters quasi-2D systems\cite{Glutsch} and quasi-1D systems
(such as quantum wires with finite radius or nanotubes\cite{IB06}),
in which the movement of electrons is confined in certain directions such that the transverse motion
can be averaged in comparison with the longitudinal motion. Though these systems show low-dimensional characteristics in various properties due to confinement, in the limit of $q\to0$ they eventually differ from strictly low-dimensional systems.


\section{The Wannier model in TDDFT} \label{sect:wannier}

Our minimal model and the Wannier exciton picture can be connected by considering the Fourier transform
of Eq. \parref{eqn:bkgnd:rhocvSPA}. We define an effective two-body  potential $V_\text{eh}$
via the Fourier transform of $F_{{\sss HXC}, \vect{k},\vect{k}'}^{(vc)(vc)}$:
\begin{equation}
V_\text{eh}(\vect{R},\vect{R}')=\frac{a+b}{2\pi}
\sum_{\vect{k},\vect{k}'\in\text{FBZ}}e^{-i\vect{k}\cdot\vect{R}}F_{{\sss HXC}, \vect{k},\vect{k}'}^{(vc)(vc)}e^{i\vect{k}'\cdot\vect{R}'},
\end{equation}
where $\vect R$ is a direct lattice vector. The Fourier transform of the density matrix is
\begin{equation}
\rho(\vect{R},\omega)\equiv\rho^{cv}(\vect{R},\omega)=\sum_\vect{k}e^{-i\vect{k}\cdot\vect{R}}\rho_\vect{k}^{cv}(\omega).
\end{equation}
Since Wannier excitons extends over many lattice constants, we approximate $\vect{R}$
as a continuous variable $\vect{r}$. Assuming the effective mass approximation, Eq. \parref{eqn:bkgnd:rhocvSPA} becomes
\begin{equation}
-\frac{\nabla^2}{2m_r}\rho(\vect{r},\omega)+\int\intd^3r'\;V_\text{eh}(\vect{r},\vect{r}',\omega)\rho(\vect{r}',\omega)=E\rho(\vect{r},\omega),
\label{eqn:bkgnd:TDDFTWannier}
\end{equation}
where $E$ is the excitonic binding energy, and the integration is carried out
over all space. We call Eq. \parref{eqn:bkgnd:TDDFTWannier} the
TDDFT Wannier equation, since it has the same form as Eq. \parref{eqn:bkgnd:Wannier}.
With proper choice of the approximated xc kernel, the nonlocal effective electron-hole interaction potential $V_\text{eh}$
supports bound excitonic states.

\begin{figure}
(a)
\includegraphics[height=0.8\columnwidth,angle=-90]{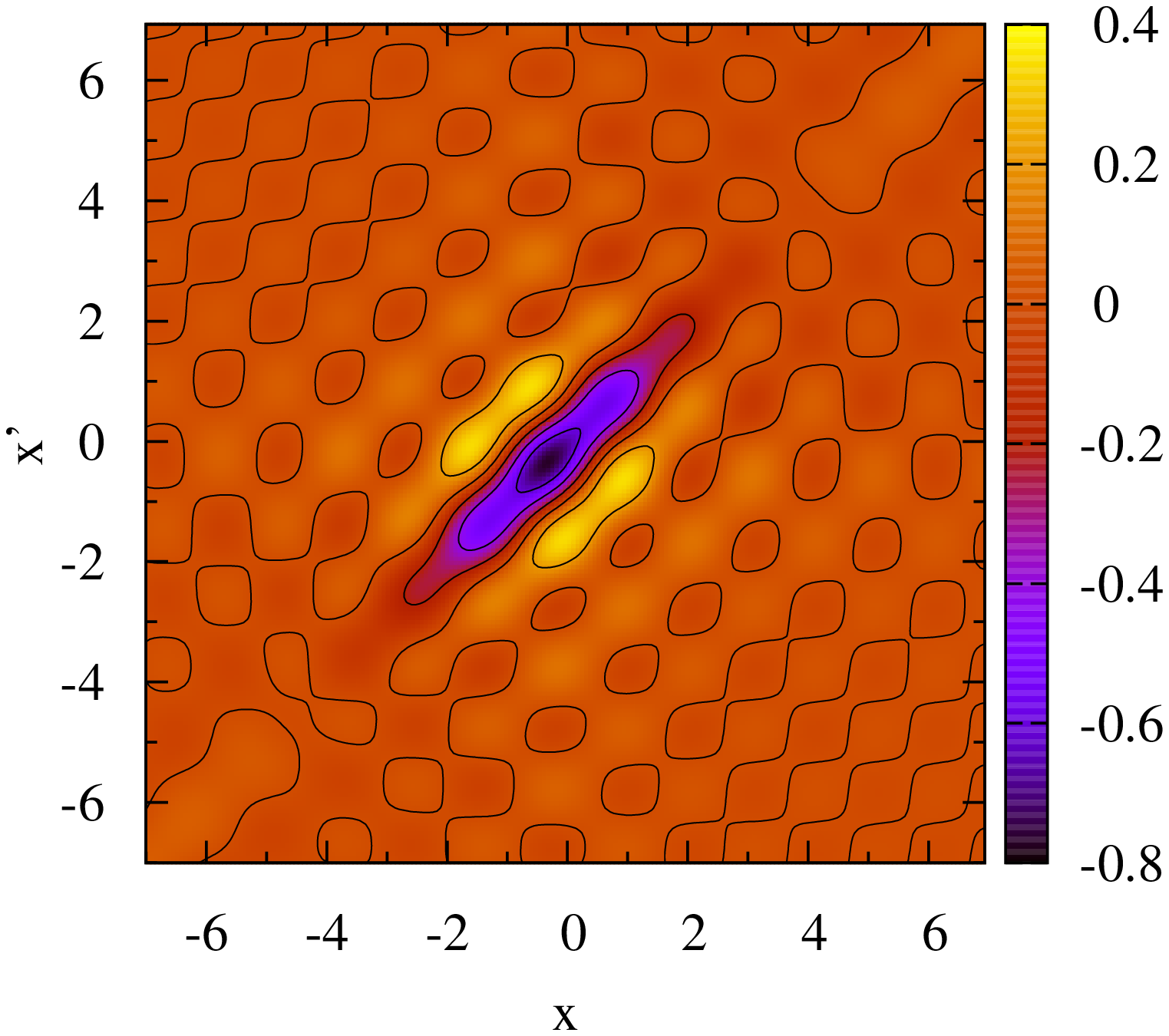}\\
(b)
\includegraphics[height=0.8\columnwidth,angle=-90]{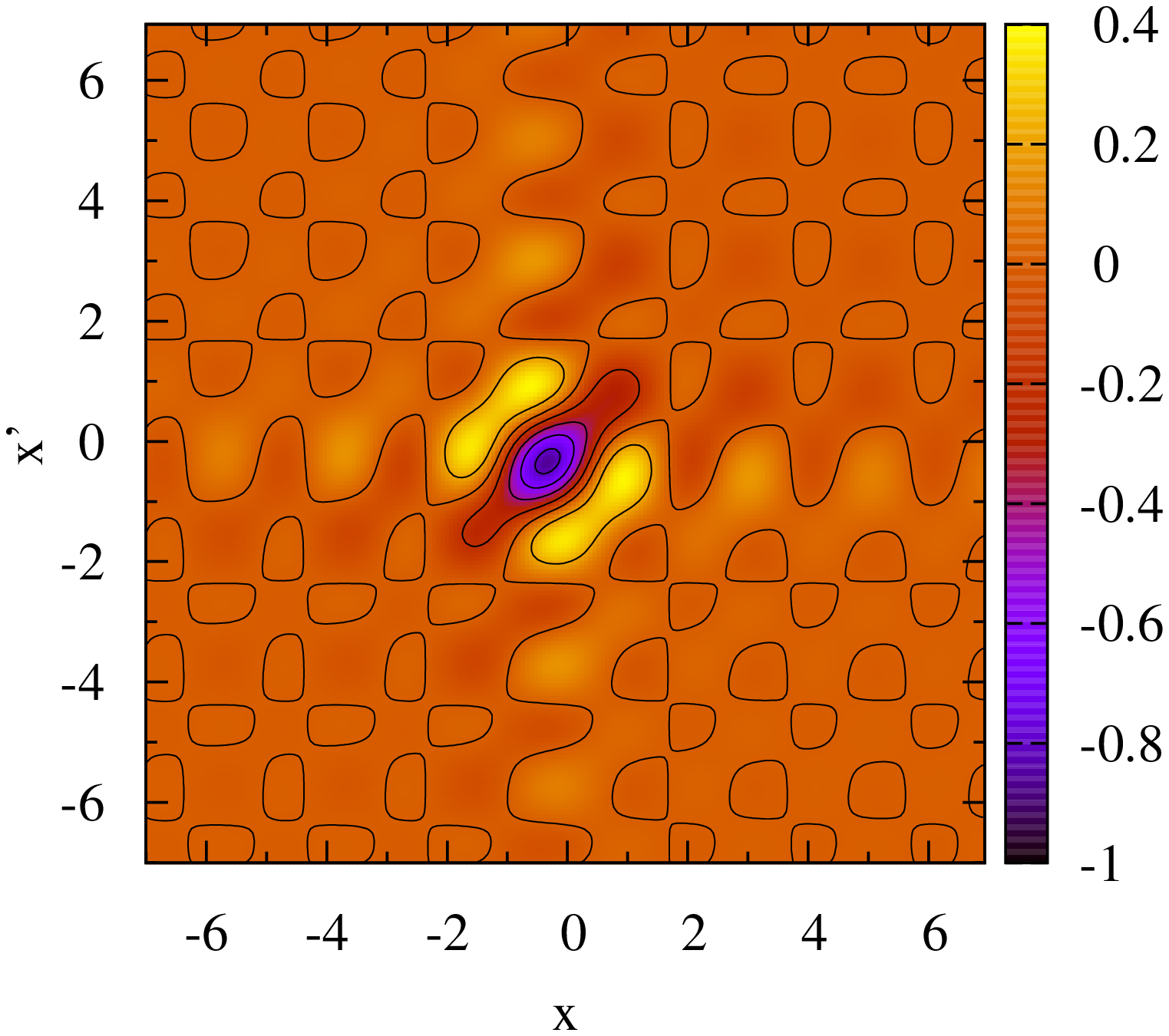}
\caption{Contour plots of $V_\text{eh}(r,r')$ for (a) BSE, and (b) TDDFT with the soft-Coulomb kernel.
The KP system is the same as in Fig. \ref{fig:results:ExcitonicEigenvect}.}
\label{fig:results:Veh}
\end{figure}

Since the BSE and the TDDFT are formally similar within the minimal model, Eq. \parref{eqn:bkgnd:rhocvSPA} can
also be applied to the BSE results. Fig. \ref{fig:results:Veh} shows the effective interaction potential
$V_\text{eh}$ for TDDFT and BSE.

The TDDFT Wannier equation provides an intuitive way of describing the effective
nonlocal electron-hole interaction, and of explaining why adiabatic TDDFT usually has fewer
excitons than BSE and the Wannier model. However, in most cases the TDDFT Wannier equation
is not suitable for quantitative use due to the approximations involved. The approximation
where we take the lattice vector $\vect{R}$ as a continuous variable assumes that the
exciton radius is much larger than the lattice constant; this works fine in most cases we
tested. But the effective mass approximation where $\omega_\vect{q}$ is approximated by $q^2/2m_r$
is only good for transitions near the band gap, thus requiring these transitions of the
noninteracting system to dominate the exciton, which is equivalent to the
exciton extending over many lattice constants. One obtains the $-\nabla^2/2m_r$ term
in Eq. \parref{eqn:bkgnd:TDDFTWannier} from $q^2/2m_r$ in the limit where
the lattice constant $a+b\to0$, and this approximation is not valid for most systems.

\begin{figure}
\includegraphics[height=0.9\columnwidth,angle=-90]{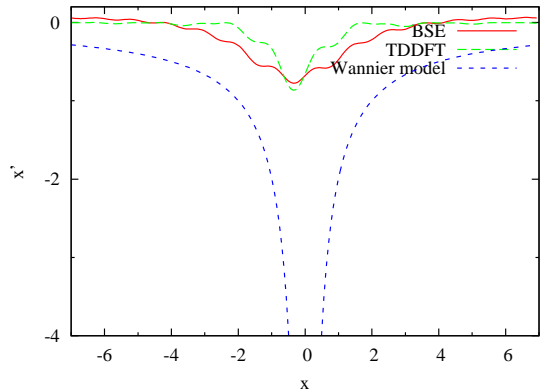}
\caption{Diagonal part $V_\text{eh}(r,r)$ of those shown in Fig. \ref{fig:results:Veh}.}
\label{fig:results:VehDiag}
\end{figure}

Although $V_\text{eh}$ is a nonlocal potential, in most cases we find that the $V_\text{eh}$'s
for both TDDFT and BSE are dominated by the diagonal part, so the exciton problem is
in analogy to one-body systems. Fig. \ref{fig:results:VehDiag} shows the diagonal part
of $V_\text{eh}$, which can be taken as the effective one-body potential. The Wannier
model in 1D has the soft-Coulomb potential, which supports an infinite number of bound
excitons (the soft-Coulomb interaction is fitted so that the binding energy
of the first exciton matches that of the BSE). We find in general that the diagonal parts of $V_\text{eh}$ for both BSE
and TDDFT are much more shallow than the soft-Coulomb potential, and the TDDFT one is more narrow
than the BSE one. Thus, BSE and TDDFT are not able to produce a complete excitonic
Rydberg series, and TDDFT in general produces fewer bound excitons than the BSE.

The TDDFT Wannier equation is not suitable for quantitative use for most of our model systems,
despite the success of the Wannier model in describing real semiconductors.\cite{UFK81, U85}
Since the approximations involved in Eq. \parref{eqn:bkgnd:TDDFTWannier} require that
the exciton radius is large compared to the lattice constant, this suggests that this discrepancy
is due to the special nature of 1D systems: namely, for similar effective masses the exciton radius in 1D is much smaller than in
3D and 2D.\cite{HK09}

\section{Conclusion}
\label{sect:app:details}

The  purpose of this paper was to construct a transparent and accessible minimalist model system that produces excitonic effects in a
non-\textit{ad hoc} fashion using TDDFT. The model, as presented here, is not intended to be a testing ground for xc kernels. Thus
despite the dimensionality restrictions for the strictly 1D system, our results in Sec. \ref{sect:results}
carry over to 3D systems.

With our minimal model, we show that adiabatic TDDFT is capable of producing bound excitons
through the local field effect even when the xc kernel is local in space, provided the strength of the kernel
is strong enough. This statement is still true in 3D; however, due to the non-vanishing head contribution
of the exact $f\xc$, we expect that the deviation of the effective interaction strength of a local $f\xc$
from the real, nonlocal $f\xc$ becomes larger than the in our strictly 1D model. In this sense the long-range kernel,
though very favorable, is not a necessary condition for excitonic effects.

We show the connection
between TDDFT and the Wannier model for excitons by deriving the TDDFT Wannier equation,
which describes a real-space system featuring a nonlocal effective electron-hole interaction. Such a connection intuitively demonstrates how
adiabatic TDDFT generally produces fewer bound excitons than BSE, and does not have a complete Rydberg
series. The eigenvectors of the excitonic excitations in the minimal model clearly show their
collective nature, which is not obvious from the Wannier model alone. Excitonic instabilities may
show up in TDDFT with approximate xc kernels, and this suggests that the TDA tends to be more reliable
for excitons than the formally exact method.

The frequency dependence of the exact xc kernel, $f\xc({\bf r},{\bf r}',\omega)$, is usually ignored. Despite the fact that adiabatic xc kernels
have met with considerable recent success in producing optical spectra of insulators and semiconductors (see the discussion in
the Introduction), they are incapable of producing excitonic Rydberg series.
Our model system gives an explanation for why this is the case. This failure of the adiabatic approximation for $f\xc$
is quite different from that which is responsible for the inability of adiabatic TDDFT to produce double excitations
in finite systems or certain classes of charge-transfer excitations.\cite{Maitra2004,Maitra2005}
This calls for continuing efforts in the search for nonadiabatic xc kernels for excitons.

\section*{Acknowledgement}
This work was supported by NSF Grant No. DMR-1005651.

\appendix
\section{The Bethe-Salpeter equation}
\label{sect:app:BSE}

Electrons and holes near the Fermi surface are well described in the quasiparticle picture. The quasiparticle
Green's function $G$ is related to that of the noninteracting system, $G_0$, through the use of the self-energy $\Sigma$:
\begin{equation}
G(12)=G_0(12)+\int\intd(34)\;G_0(13)\Sigma(34)G(42),
\end{equation}
where the arguments denote sets of space and time variables. A widely used approximation for the self-energy is
the GW approximation: \cite{H65,AG98}
\begin{equation}
\Sigma(12)=iG(12)W(12),
\label{eqn:app:BSE:GWsigma}
\end{equation}
where $W$ is the screened interaction,
\begin{equation}
W(\vect{r},\vect{r}',\omega)=\int\intd^3 r''\;\epsilon^{-1}(\vect{r},\vect{r}'',\omega)v(\vect{r}'',\vect{r}'),
\end{equation}
$v$ is the bare Coulomb interaction, and the inverse dielectric function $\epsilon^{-1}$ is obtained as
\begin{equation}
\epsilon^{-1}(\vect{r},\vect{r}',\omega)=\delta(\vect{r}-\vect{r}')+\int\intd^3 r''\;v(\vect{r},\vect{r}'')\chi(\vect{r}'',\vect{r}',\omega).
\label{eqn:app:BSE:epsiloninv}
\end{equation}
The linear response function $\chi$ can be calculated by its Lehmann representation:
\begin{equation}
\chi(\vect{r},\vect{r}',\omega)=\sum_{ij}\frac{\psi_i^*(\vect{r})\psi_j(\vect{r})\psi_i(\vect{r}')\psi_j^*(\vect{r}')}
{\omega-(E_i-E_j)+i\eta}(f_j-f_i),
\label{eqn:app:BSE:chi}
\end{equation}
where $\psi_i$ are quasiparticle states, $E_i$ are quasiparticle energies, and $f_i$ are
occupation numbers. In practice, evaluating $\chi$ through Eq. \parref{eqn:app:BSE:chi} can be quite time-consuming,
and $\chi$ is often calculated with the plasmon-pole model.\cite{HL86} However, our minimal model
is simple enough to allow us to use Eq. \parref{eqn:app:BSE:chi} directly.

The GW quasiparticle Green's function obtained from Eq. \parref{eqn:app:BSE:GWsigma} misses
important dynamical many-body effects, such as the electron-hole (excitonic) interaction.
The two-particle Green's function includes these effects.
The BSE\cite{HS80,ORR02} describes the relation between the four-point polarization function $L(1234)$ of an interacting
system and the corresponding object of the quasiparticle system:
\begin{eqnarray}
L(1234)&=&L_0(1234)+\int\intd(5678)\;L_0(1256)
\nonumber\\
&& \times K(5678)L(7834),
\label{eqn:app:BSE:L}
\end{eqnarray}
in which $L_0$ is
\begin{equation}
L_0(1234)=iG(13)G(42)
\end{equation}
and assuming the GW approximation, the kernel $K$ is
\begin{equation} \label{BSEkernel}
K(1234)=\delta(12)\delta(34)\bar{v}(13)-\delta(13)\delta(24)W(12).
\end{equation}
Here, $\bar v$ denotes the Coulomb interaction with the long-range part removed.\cite{ORR02}
In practice the BSE is often solved in the transition space, which is spanned by single-particle excitations.
A four-point function such as $L$ then becomes
\begin{multline}
L^{(ij)(mn)}(\omega)=\int\intd x_1\ldots x_4\;L(\vect{r}_1\vect{r}_2\vect{r}_3\vect{r}_4;\omega)\\
\times\phi_i(\vect{r}_1)\phi_j^*(\vect{r}_2)\phi_m^*(\vect{r}_3)\phi_n(\vect{r}_4),
\end{multline}
where the $\phi$'s can be any complete basis set. Eq. \parref{eqn:app:BSE:L} in the transition space becomes
\begin{equation}
L^{(ij)(mn)}(\omega)=[H_\text{exc}(\omega)-I\omega]^{-1}_{(ij)(mn)}(f_m-f_n),
\end{equation}
where the excitonic Hamiltonian matrix is
\begin{eqnarray}
H_\text{exc}^{(ij)(mn)}(\omega)&=&(E_j-E_i-\omega)\delta_{im}\delta_{jn} \nonumber \\
&+&(f_i-f_j){K}^{(ij)(mn)}(\omega).
\end{eqnarray}
We make the adiabatic approximation for $H_\text{exc}^{(ij)(mn)}$ and arrive at the following eigenvalue problem:
\begin{equation}
\sum_{mn}H_\text{exc}^{(ij)(mn)}A_\lambda^{(mn)}=\omega^\text{exc}_\lambda A_\lambda^{(ij)},
\end{equation}
and $L^{(ij)(mn)}$ can be expressed in terms of these eigenvectors by
\begin{equation}
L^{(ij)(mn)}(\omega)=\sum_\lambda\frac{A_\lambda^{(ij)}A_\lambda^{(mn)*}}{\omega^\text{exc}_\lambda-\omega} .
\end{equation}
Only the transitions between the valence and conduction bands contribute. Within our two-band model,
the excitonic Hamiltonian has the following block matrix form:
\begin{equation}
H_\text{exc}=\left(\begin{array}{cc} E_c-E_v +K^{(vc)(vc)} & K^{(vc)(cv)}\\
-K^{(vc)(cv)*} & E_v-E_c-K^{(vc)(vc)*}\end{array}\right).
\label{eqn:app:BSE:Hexc}
\end{equation}
Ignoring the off-diagonal part in Eq. \parref{eqn:app:BSE:Hexc} is equivalent to the TDA.

As shown in Sec. \ref{sect:results}, it is possible that instabilities show up in the full
BSE results when the underlying ground-state calculation is not exact. Such instabilities in the minimal model are an
artifact originating from the assumption that the solution of the KP model constitutes the ground state of the many-body system.
However, this is not a matter of great concern in practice.

In principle, the transition space spans all possible
combinations of valence and conduction orbitals, including
nonvertical transitions connecting different Bloch wavevectors.
The kernel $K=v-W$ of the BSE in momentum space, Eq. (\ref{BSEkernel}),  has the following ingredients:
\begin{multline}
v^{(ij)(mn)}=\frac{1}{\Omega}\sum_{\vect{G}}v_\vect{G}(\vect{q})\delta_{\vect{q},\vect{k}_j-\vect{k}_i+\vect{G}_0}
\delta_{\vect{q},\vect{k}_n-\vect{k}_m+\vect{G}_0}\\
\times \matelem{j,\vect{k}_j}{e^{i(\vect{q}+\vect{G})\cdot\vect{r}}}{i,\vect{k}_i}\matelem{m,\vect{k}_m}{e^{-i(\vect{q}+\vect{G})
\cdot\vect{r}}}{n,\vect{k}_n},
\label{eqn:app:BSE:v}
\end{multline}
and
\begin{multline}
W^{(ij)(mn)}=\frac{1}{\Omega}\sum_{\vect{G},\vect{G}'}W_{\vect{G},\vect{G}'}(\vect{q})
\delta_{\vect{q},\vect{k}_j-\vect{k}_n+\vect{G}_0}\delta_{\vect{q},\vect{k}_i-\vect{k}_m+\vect{G}_0}\\
\times \matelem{j,\vect{k}_j}{e^{i(\vect{q}+\vect{G})\cdot\vect{r}}}{n,\vect{k}_n}\matelem{m,\vect{k}_m}{e^{-i(\vect{q}+\vect{G}')
\cdot\vect{r}}}{i,\vect{k}_i}.
\label{eqn:app:BSE:w}
\end{multline}
Only the excitations with the same momentum transfer $\bf q$ are
coupled due to the $\delta$ functions in Eq. \parref{eqn:app:BSE:v} and \parref{eqn:app:BSE:w},
so we only need to include vertical transitions in the calculations for optical properties.


\end{document}